\documentclass[aps,prd,superscriptaddress,twoside,showpacs,
nofootinbib,10pt,floatfix,showpacs,twocolumn]{revtex4-1}

\usepackage{bm}
\usepackage{tikz}
\usepackage{amsmath}
\usepackage{multirow}
\usepackage{gensymb}

\begin{document}

\title{Pentaquarks with the $qqs\bar{Q}Q$ configuration in the Chiral Quark Model}


\author{Qi Zhang}
\affiliation{
Department of Physics, Nanjing Normal University, Nanjing 210023, PR China
}
\author{Bing-Ran He}
\email[E-mail: ]{hebingran@njnu.edu.cn (Corresponding author)}
\affiliation{
Department of Physics, Nanjing Normal University, Nanjing 210023, PR China
}
\author{Jia-Lun Ping}
\email[E-mail: ]{jlping@njnu.edu.cn (Corresponding author)}
\affiliation{
Department of Physics, Nanjing Normal University, Nanjing 210023, PR China
}


\date{\today}

\begin{abstract} 
    We study the five-quark system composed of $qqs\bar{Q}Q$ configuration ($q = u$ or $d$, $Q=b$ or $c$), in the framework of the chiral quark model. In consequence, a series of bound states with heavy flavors are predicted by precise five-body dynamical calculations. We found that taking color-octet structure into consideration always provides more bounding energy than color-singlet structure, and the more heavier quark prevents, the easier to form the bound states. We suggest $qqs\bar{b}b$ configuration is a compact $\bar{b}b$-pair surrounded by three other quarks, while $qqs\bar{b}c$, $qqs\bar{c}b$ and $qqs\bar{c}c$ configurations are molecular states.
\end{abstract}


\maketitle

\section{\label{sec:level1}Introduction}
    In 1964, Quark model was proposed by M.Gell-Mann and G.Zweig respectively, which has made great achievements in describing hadron spectrum and hadron properties~\cite{GellMann:1964nj,Zweig:1964jf}.  
In conventional quark models, baryons and mesons are described as simple quark-antiquark ($q\bar{q}$) and 3-quark ($qqq$) configurations~\cite{Nakamura:2010zzi}. While the classical quark model is very successful in explaining the properties of the spatial ground states of the flavor SU(3) vector meson nonet, baryon octet, and decuplet, it fails badly for the spatial excited states in both meson and baryon sectors~\cite{Guo:2017jvc}.  
The structure of hadrons can be glueballs, hybrids and multiquark states in quantum chromodynamics (QCD), which is the basic theory of strong interactions. Actually, Gell-Mann also pointed out that the quark model does not prevent the existence of multiquark states.  
Researches for exotic hadrons have been done by many theoretical workers, some baryon resonances are proposed to be hadronic molecules~\cite{Kaiser:1995cy,Oller:2000fj,Inoue:2001ip,GarciaRecio:2003ks} or states with large pentaquark components~\cite{Helminen:2000jb,Zhu:2004xa,Liu:2005pm,Bijker:2009up,An:2009uv,Liu:2019zoy}. 
    
    Recently, the LHCb Collaboration observed several hidden-charm pentaquark states, $P_c^+(4312)$, $P_c^+(4380)$, $P_c^+(4440)$, $P_c^+(4457)$ in the $J/\psi p$ invariant mass distribution of the $\Lambda_b^0 \rightarrow J/\psi K^-p$ decay~\cite{Aaij:2015tga,Aaij:2019vzc}. This has set off an upsurge in investigations of pentaquark states~\cite{He:2015cea,Liu:2015fea,Scoccola:2015nia,Wang:2015ava,Shen:2016tzq,Zhou:2018bkn,Li:2018vhp,An:2019idk,Zhang:2020erj}. In fact, the hidden-charm pentaquark states have been predicted by J.J.Wu \emph{et al.} with the coupled-channel unitary approach a few years before the LHCb discovery~\cite{Wu:2010jy}. Afterwards, they extend the study to the hidden-bottom sector, and some super-heavy $N^*(uud\bar{b}b)$ and $\Lambda^*(uds\bar{b}b)$ resonances are predicted to exist, with a mass around 11 GeV and width smaller than 10 MeV~\cite{Wu:2010rv}. In Ref.~\cite{Yang:2018oqd,Shimizu:2016rrd}, the possible hidden-bottom pentaquark states were also predicted to exist. Besides the theoretical works, the LHCb Collaboration has tried to find pentaquark states with a single bottom quark~\cite{Aaij:2017jgf}. Therefore, explorations of pentaquarks in the heavy flavor sector should be expected in the future. 
Based on the above experimental and theoretical works, we investigate the five-quark systems with heavy flavors and strangeness $S = -1$, composed of $qqs\bar{Q}Q$ configuration ($q = u$ or $d$, $Q=b$ or $c$), in the framework of the chiral quark model. 
We consider the possible quantum numbers as $IJ^P = 0(\frac{1}{2})^-$, $IJ^P = 0(\frac{3}{2})^-$, $IJ^P = 1(\frac{1}{2})^-$, $IJ^P = 1(\frac{3}{2})^-$, $IJ^P = 1(\frac{5}{2})^-$ and both color-singlet and color-octet structures. 
We perform a precise five-body dynamical calculation to search for possible bound states systematically. 
We found that: (1), taking color-octet structure into consideration always provide more bounding energy than color-singlet structure, which is consisted with Refs.~\cite{Yang:2018oqd,1797080}; (2), the more heavier quark prevents, the easier to form the bound states, i.e., the bounding energy $E_B$ satisfies $|E_B(qqs\bar{b}b)|>|E_B(qqs\bar{b}c)| \simeq |E_B(qqs\bar{c}b)|>|E_B(qqs\bar{c}c)|$.
The distances between two quarks and expectations of potential energy are calculated to explore the structures and attraction mechanism, respectively. 
We found that $qqs\bar{b}b$ is a compact $\bar{b}b$-pair which is surrounded by three other quarks~\cite{Yang:2018oqd}, while $qqs\bar{b}c$, $qqs\bar{c}b$ and $qqs\bar{c}c$ are molecular states. 
    
    The structure of this paper is organized as follows. In Section.~\ref{sec:level2}, details about the chiral quark model, methods for the multiquark system and quark model parameters are introduced. In Section.~\ref{sec:level3}, the numerical results with analysis and discussion are presented. In Section.~\ref{sec:level4}, we give a brief summary of this work. 

\section{\label{sec:level2}Theoretical Framework}
\subsection{The chiral quark model}
    Quark Model is one of the common tools to study multiquark states observed by experiments. This work is discussed in the framework of the Chiral Quark Model (ChQM), which has made lots of achievements in describing hadron spectrum and hadron-hadron interaction~\cite{Obukhovsky:1990tx,Fernandez:1993hx,Yu:1995ag,Valcarce:1995dm,Fernandez:2019ses}. The specific introduction of the model can be found in Ref.~\cite{Vijande:2004he}. 
The Hamiltonian for multiquark system is given as follows,
    \begin{eqnarray}
        && H = \sum_{i=1}^n \left( m_i + \frac{\bm{p}_i^2}{2m_i} \right) - T_{cm} + \sum_{j>i=1}^n V_{ij}, \\
        && V_{ij} = V_{ij}^C + V_{ij}^G + V_{ij}^{\pi} + V_{ij}^K + V_{ij}^{\eta} + V_{ij}^{\sigma}.
    \end{eqnarray}
    Where $m_i$ is the mass of constituent quarks and $T_{cm}$ is the kinetic energy of the center of mass motion.  $V_{ij}^{\pi}, V_{ij}^K, V_{ij}^{\eta}$ are the Goldstone boson exchange interactions between light quarks, to reconstruct the chiral symmetry in the QCD Lagrangian. $V_{ij}^C$ and $V_{ij}^G$  are phenomenology confinement potential and one-gluon-exchange potential, respectively, which are flavor blindness. Scalar meson potential $V_{ij}^{\sigma}$ is expected to exchange between $u(\bar{u})$ and $d(\bar{d})$. Because we are interested in the lowest-lying states of multiquark system, only the central part of the interaction is given below,
    \begin{widetext}
    \begin{eqnarray}
        && V_{ij}^C = \left( \bm{\lambda}_{i}^c \cdot \bm{\lambda}_{j}^c \right)
                      \left[ -a_c \left( 1 - e^{-\mu_c r_{ij}} \right) + \Delta \right], \\
        && V_{ij}^G = \frac{1}{4} \alpha_s \left( \bm{\lambda}_{i}^c \cdot \bm{\lambda}_{j}^c \right)
                      \left[ \frac{1}{r_{ij}} - \frac{1}{6m_im_j} \frac{e^{-\frac{r_{ij}}{r_0(\mu_{ij})}}}{r_{ij}r_0^2(\mu_{ij})}
                      \bm{\sigma}_i \cdot \bm{\sigma}_j \right], \\
        && V_{ij}^{\sigma} = -\frac{g_{ch}^2}{4\pi} \frac{\Lambda_{\sigma}^2}{\Lambda_{\sigma}^2 - m_{\sigma}^2} m_{\sigma}
                             \left[ Y(m_{\sigma}r_{ij}) - \frac{\Lambda_{\sigma}}{m_{\sigma}} Y(\Lambda_{\sigma}r_{ij}) \right], \\
        && V_{ij}^{\pi} = \frac{g_{ch}^2}{4\pi} \frac{m_{\pi}^2}{12m_im_j} \frac{\Lambda_{\pi}^2}{\Lambda_{\pi}^2 - m_{\pi}^2} m_{\pi}
                          \left[ Y(m_{\pi}r_{ij}) - \frac{\Lambda_{\pi}^3}{m_{\pi}^3} Y(\Lambda_{\pi}r_{ij}) \right] \bm{\sigma}_i\cdot\bm{\sigma}_j \sum_{a=1}^3 \lambda_{i}^a \lambda_{j}^a, \\
        && V_{ij}^{K} = \frac{g_{ch}^2}{4\pi} \frac{m_{K}^2}{12m_im_j} \frac{\Lambda_{K}^2}{\Lambda_{K}^2 - m_{K}^2} m_{K}
                        \left[ Y(m_{K}r_{ij}) - \frac{\Lambda_{K}^3}{m_{K}^3} Y(\Lambda_{K}r_{ij}) \right] \bm{\sigma}_i\cdot\bm{\sigma}_j
                        \sum_{a=4}^7 \lambda_{i}^a \lambda_{j}^a, \\
        && V_{ij}^{\eta} = \frac{g_{ch}^2}{4\pi} \frac{m_{\eta}^2}{12m_im_j} \frac{\Lambda_{\eta}^2}{\Lambda_{\eta}^2 - m_{\eta}^2} m_{\eta}
                           \left[ Y(m_{\eta}r_{ij}) - \frac{\Lambda_{\eta}^3}{m_{\eta}^3} Y(\Lambda_{\eta}r_{ij}) \right] \bm{\sigma}_i\cdot\bm{\sigma}_j
                           \left( \lambda_{i}^8\lambda_{j}^8 \cos{\theta_p} - \lambda_{i}^0\lambda_{j}^0 \sin{\theta_p} \right).
    \end{eqnarray}
    \end{widetext}
    Where $\bm{\lambda}$ and $\bm{\lambda}^c$ are $SU(3)$ Gell-Mann matrices of flavor and color, respectively. $\mu_{ij}$ is the reduced mass of two interacting quarks, and $r_0(\mu_{ij}) = \hat{r}_0 / \mu_{ij}$ is a regulator that depends on $\mu_{ij}$. $\bm{\sigma}$ is the $SU(2)$ Pauli matrix, and the chiral coupling constant $g_{ch}$ is determined from the pion-nucleon coupling constant. $Y(x) = e^{-x}/x$ is the standard Yukawa functions. 
The running property of the one-gluon-exchange coupling constant $\alpha_s$ is given as~\cite{Segovia:2013wma},
    \begin{equation}
        \alpha_s(\mu_{ij}) = \frac{\alpha_0}{\ln \left[ \left( \mu_{ij}^{2} + \mu_0^2 \right) / \Lambda_{0}^2\right]}.
    \end{equation}
    
\subsection{Methods for the multiquark system}
    \begin{figure}[b]
    \begin{tikzpicture}
            \filldraw [fill=white,thick] (0,0) circle [radius=8pt]  ;
            \filldraw [fill=white,thick] (3,0) circle [radius=8pt] ;
            \filldraw [fill=white,thick] (60:3) circle [radius=8pt] ;
            \filldraw [thick] (5,3) circle [radius=8pt] ;
            \filldraw [fill=white,thick] (6,1) circle [radius=8pt]  ;

            \fill [fill=black] (0,0) circle [radius=1pt] ;
            \fill [fill=black] (3,0) circle [radius=1pt] ;
            \fill [fill=black] (60:3) circle [radius=1pt] ;
            \fill [fill=black] (5,3) circle [radius=1pt] ;
            \fill [fill=black] (6,1) circle [radius=1pt] ;

            \draw[thick] (3,0) -- node[below]{$\bm{r}_{12}$} (0,0) ;
            \draw[thick] (1.5,0) -- node[left]{$\bm{r}_{12,3}$} (60:3) ;
            \draw[thick] (5,3) -- node[right]{$\bm{r}_{45}$} (6,1) ;
            \draw[thick] (1.5,1) -- node[above]{$\bm{r}_{123,45}$} (5.5,2) ;

            \node[below left = 8pt] at (0,0) {$N_1$} ;
            \node[below right = 8pt] at (3,0) {$N_2$} ;
            \node[above = 10pt] at (60:3) {$N_3$} ;
            \node[above left = 8pt] at (5,3) {$N_4$} ;
            \node[below right = 8pt] at (6,1) {$N_5$} ;
        \end{tikzpicture}
        \caption{\label{fig:epsart}The molecular structure of the five-quark system. The hollow circles stand for quark and the black circle stands for antiquark.}
    \end{figure}
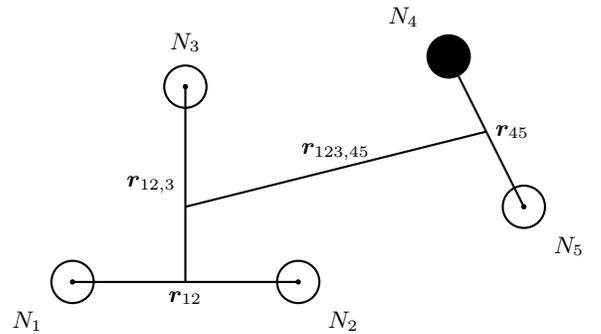
    As shown in Fig.~\ref{fig:epsart}, the orbital wave function of the five-quark system can be obtained by coupling the orbital wave function of each relative motion, and is written as follows,
    \begin{equation}
         \psi_{LM_L} = [ [ [ \psi_{l_1}(\bm{r}_{12}) \psi_{l_2}(\bm{r}_{12,3}) ]_l
                               \psi_{l_3}(\bm{r}_{45}) ]_{l'} \psi_{l_4}(\bm{r}_{123,45}) ]_{LM_L},
    \end{equation}
where
\begin{eqnarray}
\bm{r}_{12} &=& \bm{r}_{1} - \bm{r}_{2}\,, \nonumber\\
        \bm{r}_{12,3} &=& \frac{m_1\bm{r}_{1}+m_2\bm{r}_{2}}{m_1+m_2} - \bm{r}_{3}\,, \nonumber\\
        \bm{r}_{45} &=& \bm{r}_{4} - \bm{r}_{5}\,, \nonumber\\
        \bm{r}_{123,45} &=& \frac{m_1\bm{r}_{1} + m_2\bm{r}_{2} + m_3\bm{r}_{3}}
                          {m_1+m_2+m_3} - \frac{m_4\bm{r}_{4}+m_5\bm{r}_{5}}{m_4+m_5}\,.\,\,
\end{eqnarray}  
Here $\psi_{l_1}(\bm{r}_{12}), \psi_{l_2}(\bm{r}_{12,3}), \psi_{l_3}(\bm{r}_{45})$ are the orbital wave functions of relative motions in each cluster respectively, and $\psi_{l_4}(\bm{r}_{123,45})$ is the orbital wave function between two clusters, square brackets represents the angular momentum coupling. 
In the present work, the orbital wave functions of the system are determined by the gaussian expansion method (GEM)~\cite{Hiyama:2003cu}. In GEM, the orbital wave function is written as the radial part and spherical harmonics, and the radial wave function is expanded by gaussians,
    \begin{eqnarray}
        && \psi_{lm}(\bm{r}) = \sum_{n=1}^{n_{max}} c_n \psi_{nlm}^G(\bm{r}), \\
        && \psi_{nlm}^G(\bm{r}) = N_{nl} r^l e^{-\nu_{n}r^2} Y_{lm}(\hat{\bm{r}}), \\
        && N_{nl} = \left( \frac{2^{l+2}(2\nu_n)^{l+\frac{3}{2}}}{\sqrt{\pi}(2l+1)!!} \right)^{\frac{1}{2}}, \\
        && \nu_n = \frac{1}{r_n^2}, r_n = r_{min}a^{n-1}, a = \left( \frac{r_{max}}{r_{min}}
                   \right)^{\frac{1}{n_{max}-1}}.
    \end{eqnarray}
    Where $c_n$ are expansion coefficients obtained by solving the Schrodinger equation; $N_{nl}$ are normalization constants; The Gaussian size $\nu_n$ are taken as the geometric progression to simplify the calculation of Hamiltonian matrix elements. For the present study, $n_{max}=7$ leads to a converge result. 
The ground state energy and the radial excitation energy can be obtained by diagonalizing the Hamiltonian matrix.

    The spin wave functions of 3-quark and 2-quark clusters are,
    \begin{eqnarray}
        && \chi_{\frac{3}{2},\frac{3}{2}}^{\sigma}(3) = \alpha\alpha\alpha, \nonumber \\
        && \chi_{\frac{3}{2},\frac{1}{2}}^{\sigma}(3) = \sqrt{\frac{1}{3}}
                                                        (\alpha\alpha\beta+\alpha\beta\alpha+\beta\alpha\alpha), \nonumber \\
        && \chi_{\frac{3}{2},-\frac{1}{2}}^{\sigma}(3) = \sqrt{\frac{1}{3}}
                                                           (\alpha\beta\beta+\beta\alpha\beta+\beta\beta\alpha), \nonumber \\
        && \chi_{\frac{3}{2},-\frac{3}{2}}^{\sigma}(3) = \beta\beta\beta, 
\end{eqnarray}

\begin{eqnarray}
        && \chi_{\frac{1}{2},\frac{1}{2}}^{\sigma1}(3) = \sqrt{\frac{1}{6}}
                                                           (2\alpha\alpha\beta-\alpha\beta\alpha-\beta\alpha\alpha), \nonumber \\
        && \chi_{\frac{1}{2},\frac{1}{2}}^{\sigma2}(3) = \sqrt{\frac{1}{2}}
                                                           (\alpha\beta\alpha-\beta\alpha\alpha), \nonumber \\
        && \chi_{\frac{1}{2},-\frac{1}{2}}^{\sigma1}(3) = \sqrt{\frac{1}{6}}
                                                            (\alpha\beta\beta+\beta\alpha\beta-2\beta\beta\alpha), \nonumber \\
        && \chi_{\frac{1}{2},-\frac{1}{2}}^{\sigma2}(3) = \sqrt{\frac{1}{2}}
                                                            (\alpha\beta\beta-\beta\alpha\beta),     
\end{eqnarray}

\begin{eqnarray}
        && \chi_{1,1}^{\sigma}(2) = \alpha\alpha, \nonumber \\
        && \chi_{1,0}^{\sigma}(2) = \sqrt{\frac{1}{2}}(\alpha\beta+\beta\alpha), \nonumber \\
        && \chi_{1,-1}^{\sigma}(2) = \beta\beta, \nonumber \\
        && \chi_{0,0}^{\sigma}(2) = \sqrt{\frac{1}{2}}(\alpha\beta-\beta\alpha).
    \end{eqnarray}
Where $\alpha$ and $\beta$ denote spin-up and spin-down states, respectively.

    The spin wave functions for the five-quark system are constructed by coupling the spin wave functions of each cluster using Clebsch-Gordan Coefficients,
    \begin{eqnarray}
         \chi_{\frac{1}{2},\frac{1}{2}}^{\sigma1}(5) = && \sqrt{\frac{1}{6}}\chi_{\frac{3}{2},-\frac{1}{2}}^{\sigma}(3)\chi_{1,1}^{\sigma}(2) -
                                                           \sqrt{\frac{1}{3}}\chi_{\frac{3}{2},\frac{1}{2}}^{\sigma}(3)\chi_{1,0}^{\sigma}(2)  \nonumber \\
                                                       &&  +\sqrt{\frac{1}{2}}\chi_{\frac{3}{2},\frac{3}{2}}^{\sigma}(3)\chi_{1,-1}^{\sigma}(2), \nonumber \\
         \chi_{\frac{1}{2},\frac{1}{2}}^{\sigma2}(5) = && \sqrt{\frac{1}{3}}\chi_{\frac{1}{2},\frac{1}{2}}^{\sigma1}(3)\chi_{1,0}^{\sigma}(2) -
                                                           \sqrt{\frac{2}{3}}\chi_{\frac{1}{2},-\frac{1}{2}}^{\sigma1}(3)\chi_{1,1}^{\sigma}(2), \nonumber \\
         \chi_{\frac{1}{2},\frac{1}{2}}^{\sigma3}(5) = && \sqrt{\frac{1}{3}}\chi_{\frac{1}{2},\frac{1}{2}}^{\sigma2}(3)\chi_{1,0}^{\sigma}(2) -
                                                           \sqrt{\frac{2}{3}}\chi_{\frac{1}{2},-\frac{1}{2}}^{\sigma2}(3)\chi_{1,1}^{\sigma}(2), \nonumber \\
         \chi_{\frac{1}{2},\frac{1}{2}}^{\sigma4}(5) = && \chi_{\frac{1}{2},\frac{1}{2}}^{\sigma1}(3)\chi_{0,0}^{\sigma}(2), \nonumber \\
         \chi_{\frac{1}{2},\frac{1}{2}}^{\sigma5}(5) = && \chi_{\frac{1}{2},\frac{1}{2}}^{\sigma2}(3)\chi_{0,0}^{\sigma}(2), 
\end{eqnarray}

\begin{eqnarray}
         \chi_{\frac{3}{2},\frac{3}{2}}^{\sigma1}(5) = && \sqrt{\frac{3}{5}}\chi_{\frac{3}{2},\frac{3}{2}}^{\sigma}(3)\chi_{1,0}^{\sigma}(2) -
                                                           \sqrt{\frac{2}{5}}\chi_{\frac{3}{2},\frac{1}{2}}^{\sigma}(3)\chi_{1,1}^{\sigma}(2), \nonumber \\
         \chi_{\frac{3}{2},\frac{3}{2}}^{\sigma2}(5) = && \chi_{\frac{3}{2},\frac{3}{2}}^{\sigma}(3)\chi_{0,0}^{\sigma}(2), \nonumber \\
         \chi_{\frac{3}{2},\frac{3}{2}}^{\sigma3}(5) = && \chi_{\frac{1}{2},\frac{1}{2}}^{\sigma1}(3)\chi_{1,1}^{\sigma}(2), \nonumber \\
         \chi_{\frac{3}{2},\frac{3}{2}}^{\sigma4}(5) = && \chi_{\frac{1}{2},\frac{1}{2}}^{\sigma2}(3)\chi_{1,1}^{\sigma}(2), \nonumber \\
         \chi_{\frac{5}{2},\frac{5}{2}}^{\sigma1}(5) = && \chi_{\frac{3}{2},\frac{3}{2}}^{\sigma}(3)\chi_{1,1}^{\sigma}(2).
    \end{eqnarray} 
    
    In a similar way, the flavor wave functions of the five-quark system can be obtained as follows (take quark content $qqs\bar{b}b$ as an example),
    \begin{eqnarray}
         \chi_{0,0}^{f1}(5) &=& \chi_{0,0}^{f1}(3)\chi_{0,0}^{f1}(2), \nonumber \\
         \chi_{0,0}^{f2}(5) &=& \chi_{0,0}^{f2}(3)\chi_{0,0}^{f2}(2), \nonumber \\
         \chi_{0,0}^{f3}(5) &=& \sqrt{\frac{1}{2}} \chi_{\frac{1}{2},\frac{1}{2}}^{f}(3)\chi_{\frac{1}{2},-\frac{1}{2}}^{f}(2) \nonumber\\
         &&-
                                  \sqrt{\frac{1}{2}} \chi_{\frac{1}{2},-\frac{1}{2}}^{f}(3)\chi_{\frac{1}{2},\frac{1}{2}}^{f}(2), \end{eqnarray}
                                  
\begin{eqnarray}
        && \chi_{1,1}^{f1}(5) = \chi_{\frac{1}{2},\frac{1}{2}}^{f}(3)\chi_{\frac{1}{2},\frac{1}{2}}^{f}(2), \nonumber \\
        && \chi_{1,1}^{f2}(5) = \chi_{1,1}^{f1}(3)\chi_{0,0}^{f1}(2), \nonumber \\
        && \chi_{1,1}^{f3}(5) = \chi_{1,1}^{f2}(3)\chi_{0,0}^{f2}(2).
    \end{eqnarray}
    Where $\chi_I^f(3), \chi_I^f(2)$ are the flavor wave functions of each sub-cluster based on the SU(2) flavor symmetry,
    \begin{eqnarray}
        && \chi_{\frac{1}{2},\frac{1}{2}}^{f}(3) = usb, \quad \chi_{\frac{1}{2},-\frac{1}{2}}^{f}(3) = dsb, \nonumber \\
        && \chi_{1,1}^{f1}(3) = uus, \quad \chi_{1,1}^{f2}(3) = uub, \nonumber \\
        && \chi_{1,0}^{f1}(3) = \sqrt{\frac{1}{2}}(ud + du)s, \nonumber \\
        && \chi_{1,0}^{f2}(3) = \sqrt{\frac{1}{2}}(ud + du)b, \nonumber \\
        && \chi_{1,-1}^{f1}(3) = dds, \quad \chi_{1,-1}^{f2}(3) = ddb, \nonumber \\
        && \chi_{0,0}^{f1}(3) = \sqrt{\frac{1}{2}}(ud - du)s, \nonumber \\
        && \chi_{0,0}^{f2}(3) = \sqrt{\frac{1}{2}}(ud - du)b,\end{eqnarray}

\begin{eqnarray}
        && \chi_{0,0}^{f1}(2) = \bar{b}b, \quad \chi_{0,0}^{f2}(2) = \bar{b}s, \nonumber \\
        && \chi_{\frac{1}{2},\frac{1}{2}}^{f}(2) = \bar{b}u, \quad \chi_{\frac{1}{2},-\frac{1}{2}}^{f}(2) = \bar{b}d.
    \end{eqnarray} 
    
    There are two kind of color structures considered for color part, i.e., the color singlet-singlet $1 \bigotimes 1$ and octet-octet $8 \bigotimes 8$. The color wave functions are given directly. There are two kinds of the color-octet wave function $\chi_2^c$ and $\chi_3^c$, represent symmetry and antisymmetry between $N_1$ and $N_2$ in the 3-quark cluster (as shown in Fig.~\ref{fig:epsart}), respectively,
    \begin{widetext}
    \begin{eqnarray}
        \chi_1^c  = \sqrt{\frac{1}{6}} ( &&  rgb - grb + brg - rbg + gbr - bgr)
                    \sqrt{\frac{1}{3}} (\bar{r}r + \bar{g}g + \bar{b}b), \\
        \chi_2^c  = \sqrt{\frac{1}{8}} \big{[} && \sqrt{\frac{1}{6}}(2rrg-rgr-grr)\bar{r}b + \sqrt{\frac{1}{6}}(rgg+grg-2ggr)\bar{g}b - 
         \sqrt{\frac{1}{6}}(2rrb-rbr-brr)\bar{r}g - \sqrt{\frac{1}{6}}(rbb+brb-2bbr)\bar{b}g  \nonumber \\
                                                  && + \sqrt{\frac{1}{6}}(2ggb-gbg-bgg)\bar{g}r + \sqrt{\frac{1}{6}}(gbb+bgb-2bbg)\bar{b}r + 
\sqrt{\frac{1}{24}}(rbg-gbr+brg-bgr)(2\bar{b}b-\bar{r}r-\bar{g}g)  \nonumber \\
                                                  && + \sqrt{\frac{1}{24}}(2rgb-rbg+2grb-gbr-brg-bgr)(\bar{r}r-\bar{g}g) \big{]}, \\
        \chi_3^c  = \sqrt{\frac{1}{8}} \big{[} && \sqrt{\frac{1}{2}}(rgr-grr)\bar{r}b + \sqrt{\frac{1}{2}}(rgg-grg)\bar{g}b - 
         \sqrt{\frac{1}{2}}(rbr-brr)\bar{r}g - \sqrt{\frac{1}{2}}(rbb-brb)\bar{b}g  \nonumber \\
                                                  &&+ \sqrt{\frac{1}{2}}(gbg-bgg)\bar{g}r + \sqrt{\frac{1}{2}}(gbb-bgb)\bar{b}r + 
                                                   \sqrt{\frac{1}{8}}(rbg+gbr-brg-bgr)(\bar{r}r-\bar{g}g)  \nonumber \\
                                                  &&+ \sqrt{\frac{1}{72}}(2rgb+rbg-2grb-gbr-brg+bgr)(2\bar{b}b-\bar{g}g-\bar{r}r) \big{]}.
    \end{eqnarray}
    \end{widetext} 
    
    Finally, the total channel wave function for the five-quark system is a product of orbit, spin, flavor and color wave functions, 
    \begin{equation}
        \Psi_{JM_JIM_I}^{ijk} = \mathcal{A} \left[ \left[ \psi_L \chi_S^{\sigma i} \right]_{JM_J} \chi_I^{fj} \chi_k^c
                                       \right].
    \end{equation}
    Here, $\mathcal{A}$ is the antisymmetry operator, which ensures the antisymmetry of the total wave function when identical particles exchange. The eigen-energy is obtained by solving the Schrodinger equation,
    \begin{equation}
        H \Psi_{JM_JIM_I} = E \Psi_{JM_JIM_I}
    \end{equation}
    with the Rayleigh-Ritz variational principle. It is worthwhile to note that the bra vector, ket vector and central potentials in Hamiltonian may be in different Jacobian coordinates when calculating five-body matrix elements. We need to transform them into the same coordinate system and do the calculation.

\subsection{Quark model parameters}
    The model parameters of the chiral quark model are determined by fitting the meson spectrum \cite{Vijande:2004he}.  
    Some mesons and baryons involved in the present work are calculated with this set of parameters as shown in Table~\ref{tab:table2}. We notice that in Table~\ref{tab:table2}, mesons of the ground state are well described compared with the experimental value, while some baryons have much deviation.  
    In the following section, we give the energy of possible bound states by taking the sum of the experimental baryon-meson threshold and the predicted binding energy. Noting that qualitative conclusions for possible bound states should be independent of the parameters. 
    
    \begin{table}[h]
    \caption{\label{tab:table2}Meson and baryon spectrum (unit: MeV).}
    \begin{ruledtabular}
    \begin{tabular}{ c c c c c c }
            Meson & Energy & PDG & Baryon & Energy & PDG \\
            \hline
            $B$ & 5277.9 & 5279.3 & $\Sigma$ & 1341.1 & 1189.4 \\
            $B^*$ & 5318.8 & 5324.6 & $\Sigma^*$ & 1468.5 & 1382.8 \\
            $D$ & 1898.4 & 1869.6 & $\Lambda$ & 1012.9 & 1115.7 \\
            $D^*$ & 2017.3 & 2006.8 & $\Sigma_b$ & 5817.8 & 5811.3 \\
            $B_s$ & 5355.8 & 5366.9 & $\Sigma_b^*$ & 5834.6 & 5832.1 \\
            $B_s^*$ & 5400.5 & 5415.4 & $\Sigma_c$ & 2492.7 & 2454.0 \\
            $D_s$ & 1991.8 & 1968.3 & $\Sigma_c^*$ & 2536.9 & 2518.4 \\
            $D_s^*$ & 2115.7 & 2112.2 & $\Lambda_b$ & 5384.7 & 5619.6 \\
            $\eta_b$ & 9468.0 & 9400.0 & $\Lambda_c$ & 2086.3 & 2286.5 \\
            $\Upsilon$ & 9504.7 & 9460.3 & $\Xi_b$ & 5867.0 & 5791.9 \\
            $B_c$ & 6282.6 & 6274.9 & $\Xi_c$ & 2574.2 & 2467.9 \\
            $B_c^*$ & 6330.6 & - & {} & {} & {} \\
            $\eta_c$ & 2999.8 & 2983.4 & {} & {} & {} \\
            $J/\psi$ & 3096.7 & 3096.9 & {} & {} & {} \\
    \end{tabular}
    \end{ruledtabular}
    \end{table}
 
\section{\label{sec:level3}Numerical Results}
    In the present work, we try to search the possible pentaquark states of $qqs\bar{b}b$, $qqs\bar{b}c$, $qqs\bar{c}b$ and $qqs\bar{c}c$ with all possible quantum numbers $IJ^P = 0(\frac{1}{2})^-$, $IJ^P = 0(\frac{3}{2})^-$, $IJ^P = 1(\frac{1}{2})^-$, $IJ^P = 1(\frac{3}{2})^-$, $IJ^P = 1(\frac{5}{2})^-$, in the framework of the chiral quark model. 
    Because we are interested in the ground states, all the orbital angular momenta are restricted to be zero. The results of calculations of the color-singlet channel are shown in Table~\ref{tab:qqs{bbar}b} for system of $qqs\bar{b}b$, Table~\ref{tab:qqs{bbar}c} for system of $qqs\bar{b}c$, Table~\ref{tab:qqs{cbar}b} for system of $qqs\bar{c}b$, and Table~\ref{tab:qqs{cbar}c} for system of $qqs\bar{c}c$, respectively. The contributions from each term of the Hamiltonian are given in Table~\ref{tab:energy_component_qqs{bbar}b} to identify why bound states can be formed or not in some physical channels. With consideration of the color-octet structure, the channel coupling calculations of the two kind of color structures are listed in Table~\ref{tab:boundstatesof_qqs{Qbar}Q}. Furthermore, for the channel in which a bound state is found, we show the root-mean-square (RMS) distances between any two quarks in Tables~\ref{tab:RMSof_qqs{Qbar}Q_1} and \ref{tab:RMSof_qqs{Qbar}Q_18} with color-singlet structure and the coupling of two color structures, respectively, to expose structures of the pentaquark states. In the following we analyse the results in detail. 
    
    \begin{table}[h]
    \caption{\label{tab:qqs{bbar}b}The energies of pentaquark system $qqs\bar{b}b$ with the color-singlet configuration (unit: MeV).}
    \begin{ruledtabular}
    \begin{tabular}{ c c c c c c c c c }
            $IJ^P$ & Index & $[i;j;k]$ & Channel & $E$ & $E_{th}$(Theo.) & $E_B$ \\
            \hline
            $0(\frac{1}{2})^-$ & 1 & [5;1;1] & $\Lambda \eta_b$ & 10482.3 & 10480.9 & 0.0 \\
            {} & 2 & [3;1;1] & $\Lambda \Upsilon$ & 10519.0 & 10517.6 & 0.0 \\
            {} & 3 & [5;2;1] & $\Lambda_b B_s$ & 10741.1 & 10740.5 & 0.0 \\
            {} & 4 & [3;2;1] & $\Lambda_b B_s^*$ & 10785.8 & 10785.2 & 0.0 \\
            {} & 5 & [4,5;3;1] & $\Xi_b B$ & 11132.9 & 11144.9 & $-12.0$ \\
            {} & 6 & [2,3;3;1] & $\Xi_b B^*$ & 11173.4 & 11185.8 & $-12.4$ \\
            \hline
            $0(\frac{3}{2})^-$ & 1 & [4;1;1] & $\Lambda \Upsilon$ & 10519.0 & 10517.6 & 0.0 \\
            {} & 2 & [4;2;1] & $\Lambda_b B_s^*$ & 10785.8 & 10785.2 & 0.0 \\
            {} & 3 & [3,4;3;1] & $\Xi_b B^*$ & 11175.3 & 11185.8 & $-10.5$ \\
            \hline
            $1(\frac{1}{2})^-$ & 1 & [4,5;1;1] & $\Xi_b B$ & 11145.5 & 11144.9 & 0.0 \\
            {} & 2 & [2,3;1;1] & $\Xi_b B^*$ & 11186.4 & 11185.8 & 0.0 \\
            {} & 3 & [4;2;1] & $\Sigma \eta_b$ & 10810.5 & 10809.1 & 0.0 \\
            {} & 4 & [2;2;1] & $\Sigma \Upsilon$ & 10847.2 & 10845.8 & 0.0 \\
            {} & 5 & [1;2;1] & $\Sigma^* \Upsilon$ & 10974.6 & 10973.2 & 0.0 \\
            {} & 6 & [4;3;1] & $\Sigma_b B_s$ & 11174.2 & 11173.6 & 0.0 \\
            {} & 7 & [2;3;1] & $\Sigma_b B_s^*$ & 11218.8 & 11218.3 & 0.0 \\
            {} & 8 & [1;3;1] & $\Sigma_b^* B_s^*$ & 11235.6 & 11235.1 & 0.0 \\
            \hline
            $1(\frac{3}{2})^-$ & 1 & [3,4;1;1] & $\Xi_b B^*$ & 11186.4 & 11185.8 & 0.0 \\
            {} & 2 & [3;2;1] & $\Sigma \Upsilon$ & 10847.2 & 10845.8 & 0.0 \\
            {} & 3 & [2;2;1] & $\Sigma^* \eta_b$ & 10937.8 & 10936.5 & 0.0 \\
            {} & 4 & [1;2;1] & $\Sigma^* \Upsilon$ & 10974.6 & 10973.2 & 0.0 \\
            {} & 5 & [3;3;1] & $\Sigma_b B_s^*$ & 11218.8 & 11218.3 & 0.0 \\
            {} & 6 & [2;3;1] & $\Sigma_b^* B_s$ & 11190.9 & 11190.4 & 0.0 \\
            {} & 7 & [1;3;1] & $\Sigma_b^* B_s^*$ & 11235.6 & 11235.1 & 0.0 \\
            \hline
            $1(\frac{5}{2})^-$ & 1 & [1;2;1] & $\Sigma^* \Upsilon$ & 10974.6 & 10973.2 & 0.0 \\
            {} & 2 & [1;3;1] & $\Sigma_b^* B_s^*$ & 11235.6 & 11235.1 & 0.0 \\
        \end{tabular}
    \end{ruledtabular}
    \end{table}
    
    In Table~\ref{tab:qqs{bbar}b}, the third column represents the combination of spin, flavor and color degrees of freedom according to symmetry, where $i$, $j$ and $k$ are the index of spin, flavor and color wave functions, respectively. The fourth column gives the physical channels of the five-quark system. The fifth column shows the single channel eigen-energy by solving the Schrodinger equation. The sixth column is the theoretical value of noninteracting baryon-meson threshold. The values of binding energies $E_B = E - E_{th}(Theo.)$ are shown in the seventh column only if $E_B<0$. In Table~\ref{tab:qqs{bbar}b}, when isospin $I = 0$, the single channel calculation shows that no bound states are found in the channels of $\Lambda \eta_b$, $\Lambda \Upsilon$, $\Lambda_b B_s$ and $\Lambda_b B_s^*$. Their eigen-energies are a little higher than theoretical thresholds, so that no bound states can be formed. 
    However, we find the existence of bound states in the channel of $\Xi_b B$, $\Xi_b B^*(J = 1/2)$ and $\Xi_b B^*(J = 3/2)$, with the binding energies about $-12.0$ MeV, $-12.4$ MeV and $-10.5$ MeV, respectively. The result is in agreement with that of Ref.~\cite{Wu:2010rv}, in which several $\Lambda_{b\bar{b}}^*$ states dominated by the channels of $\Xi_b B$ and $\Xi_b B^*$ are predicted to exist with the coupled-channel unitary approach. With regard to states for isospin $I = 1$, the results indicate that there are no bound states formed in all possible channels. To predict more similar pentaquark states, we replace some of the $b$-quark by $c$-quark, and search for possible bound states systematically for systems of $qqs\bar{b}c$, $qqs\bar{c}b$ and $qqs\bar{c}c$, which are shown in Tables~\ref{tab:qqs{bbar}c}, \ref{tab:qqs{cbar}b} and \ref{tab:qqs{cbar}c}, respectively. 
    From Tables~\ref{tab:qqs{bbar}c} and \ref{tab:qqs{cbar}b}, we found $qqs\bar{b}c$ and $qqs\bar{c}b$ configurations have similar bound states of $qqs\bar{b}b$ configuration, but the bounding energy are about half of  $qqs\bar{b}b$ configuration when there quantum numbers are same. While for $qqs\bar{c}c$ configuration, there only exist weekly bounded states in the color-singlet channel, as shown in Table~\ref{tab:qqs{cbar}c}.  
    The physical reason for the above discussions is that, the chiral symmetry is partially restored in the heavy-light quark system~\cite{Park:2016xrw}, therefore, the more heavier $b$-quark appears, the easier to form molecular bound states in the color-singlet channel. 
    
    \begin{table}[h]
    \caption{\label{tab:qqs{bbar}c}The energies of pentaquark system $qqs\bar{b}c$ with the color-singlet configuration (unit: MeV).}
    \begin{ruledtabular}
    \begin{tabular}{ c c c c c c c }
            $IJ^P$ & Index & Channel & $E$ & $E_{th}$(Theo.) & $E_B$ \\
            \hline
            $0(\frac{1}{2})^-$ & 1 & $\Lambda B_c$ & 7296.9 & 7295.5 & 0.0 \\
            {} & 2 & $\Lambda B_c^*$ & 7344.9 & 7343.5 & 0.0 \\
            {} & 3 & $\Lambda_c B_s$ & 7443.0 & 7442.1 & 0.0 \\
            {} & 4 & $\Lambda_c B_s^*$ & 7487.7 & 7486.8 & 0.0 \\
            {} & 5 & $\Xi_c B$ & 7846.9 & 7852.1 & $-5.2$ \\
            {} & 6 & $\Xi_c B^*$ & 7887.4 & 7893.0 & $-5.6$ \\
            \hline
            $0(\frac{3}{2})^-$ & 1 & $\Lambda B_c^*$ & 7344.9 & 7343.5 & 0.0 \\
            {} & 2 & $\Lambda_c B_s^*$ & 7487.7 & 7486.8 & 0.0 \\
            {} & 3 & $\Xi_c B^*$ & 7888.7 & 7893.0 & $-4.3$ \\
            \hline
            $1(\frac{1}{2})^-$ & 1 & $\Xi_c B$ & 7852.9 & 7852.1 & 0.0 \\
            {} & 2 & $\Xi_c B^*$ & 7893.9 & 7893.0 & 0.0 \\
            {} & 3 & $\Sigma B_c$ & 7625.2 & 7623.7 & 0.0 \\
            {} & 4 & $\Sigma B_c^*$ & 7673.2 & 7671.7 & 0.0 \\
            {} & 5 & $\Sigma^* B_c^*$ & 7800.5 & 7799.1 & 0.0 \\
            {} & 6 & $\Sigma_c B_s$ & 7849.4 & 7848.5 & 0.0  \\
            {} & 7 & $\Sigma_c B_s^*$ & 7894.1 & 7893.2 & 0.0 \\
            {} & 8 & $\Sigma_c^* B_s^*$ & 7938.3 & 7937.4 & 0.0 \\
            \hline
            $1(\frac{3}{2})^-$ & 1 & $\Xi_c B^*$ & 7893.9 & 7893.0 & 0.0 \\
            {} & 2 & $\Sigma B_c^*$ & 7673.2 & 7671.7 & 0.0 \\
            {} & 3 & $\Sigma^* B_c$ & 7752.5 & 7751.1 & 0.0 \\
            {} & 4 & $\Sigma^* B_c^*$ & 7800.5 & 7799.1 & 0.0 \\
            {} & 5 & $\Sigma_c B_s^*$ & 7894.1 & 7893.2 & 0.0 \\
            {} & 6 & $\Sigma_c^* B_s$ & 7893.6 & 7892.7 & 0.0 \\
            {} & 7 & $\Sigma_c^* B_s^*$ & 7938.3 & 7937.4 & 0.0 \\
            \hline
            $1(\frac{5}{2})^-$ & 1 & $\Sigma^* B_c^*$ & 7800.5 & 7799.1 & 0.0 \\
            {} & 2 & $\Sigma_c^* B_s^*$ & 7938.3 & 7937.4 & 0.0 \\
        \end{tabular}
    \end{ruledtabular}
    \end{table}

    \begin{table}[h]
    \caption{\label{tab:qqs{cbar}b}The energies of pentaquark system $qqs\bar{c}b$ with the color-singlet configuration (unit: MeV).}
    \begin{ruledtabular}
    \begin{tabular}{ c c c c c c c }
            $IJ^P$ & Index & Channel & $E$ & $E_{th}$(Theo.) & $E_B$ \\
            \hline
            $0(\frac{1}{2})^-$ & 1 & $\Lambda B_c$ & 7296.9 & 7295.5 & 0.0 \\
            {} & 2 & $\Lambda B_c^*$ & 7344.9 & 7343.5 & 0.0 \\
            {} & 3 & $\Lambda_b D_s$ & 7377.4 & 7376.5 & 0.0 \\
            {} & 4 & $\Lambda_b D_s^*$ & 7501.3 & 7500.4 & 0.0 \\
            {} & 5 & $\Xi_b D$ & 7760.7 & 7765.4 & $-4.7$ \\
            {} & 6 & $\Xi_b D^*$ & 7880.8 & 7884.3 & $-3.5$ \\
            \hline
            $0(\frac{3}{2})^-$ & 1 & $\Lambda B_c^*$ & 7344.9 & 7343.5 & 0.0 \\
            {} & 2 & $\Lambda_b D_s^*$ & 7501.3 & 7500.4 & 0.0 \\
            {} & 3 & $\Xi_b D^*$ & 7881.7 & 7884.3 & $-2.6$ \\
            \hline
            $1(\frac{1}{2})^-$ & 1 & $\Xi_b D$ & 7766.4 & 7765.4 & 0.0 \\
            {} & 2 & $\Xi_b D^*$ & 7885.3 & 7884.3 & 0.0 \\
            {} & 3 & $\Sigma B_c$ & 7625.2 & 7623.7 & 0.0 \\
            {} & 4 & $\Sigma B_c^*$ & 7673.2 & 7671.7 & 0.0 \\
            {} & 5 & $\Sigma^* B_c^*$ & 7800.5 & 7799.1 & 0.0 \\
            {} & 6 & $\Sigma_b D_s$ & 7810.5 & 7809.6 & 0.0 \\
            {} & 7 & $\Sigma_b D_s^*$ & 7934.4 & 7933.5 & 0.0 \\
            {} & 8 & $\Sigma_b^* D_s^*$ & 7951.2 & 7950.3 & 0.0 \\
            \hline
            $1(\frac{3}{2})^-$ & 1 & $\Xi_b D^*$ & 7885.3 & 7884.3 & 0.0 \\
            {} & 2 & $\Sigma B_c^*$ & 7673.2 & 7671.7 & 0.0 \\
            {} & 3 & $\Sigma^* B_c$ & 7752.5 & 7751.1 & 0.0 \\
            {} & 4 & $\Sigma^* B_c^*$ & 7800.5 & 7799.1 & 0.0 \\
            {} & 5 & $\Sigma_b D_s^*$ & 7934.4 & 7933.5 & 0.0 \\
            {} & 6 & $\Sigma_b^* D_s$ & 7827.3 & 7826.4 & 0.0 \\
            {} & 7 & $\Sigma_b^* D_s^*$ & 7951.2 & 7950.3 & 0.0 \\
            \hline
            $1(\frac{5}{2})^-$ & 1 & $\Sigma^* B_c^*$ & 7800.5 & 7799.1 & 0.0 \\
            {} & 2 & $\Sigma_b^* D_s^*$ & 7951.2 & 7950.3 & 0.0 \\
        \end{tabular}
    \end{ruledtabular}
    \end{table}

    \begin{table}[h]
    \caption{\label{tab:qqs{cbar}c}The energies of pentaquark system $qqs\bar{c}c$ with the color-singlet configuration (unit: MeV).}
    \begin{ruledtabular}
    \begin{tabular}{ c c c c c c c }
            $IJ^P$ & Index & Channel & $E$ & $E_{th}$(Theo.) & $E_B$ \\
            \hline
            $0(\frac{1}{2})^-$ & 1 & $\Lambda \eta_c$ & 4014.4 & 4012.7 & 0.0 \\
            {} & 2 & $\Lambda J/\psi$ & 4111.3 & 4109.6 & 0.0 \\
            {} & 3 & $\Lambda_c D_s$ & 4079.4 & 4078.1 & 0.0 \\
            {} & 4 & $\Lambda_c D_s^*$ & 4203.3 & 4202.0 & 0.0 \\
            {} & 5 & $\Xi_c D$ & 4471.4 & 4472.6 & $-1.2$ \\
            {} & 6 & $\Xi_c D^*$ & 4590.7 & 4591.5 & $-0.8$ \\
            \hline
            $0(\frac{3}{2})^-$ & 1 & $\Lambda J/\psi$ & 4111.3 & 4109.6 & 0.0 \\
            {} & 2 & $\Lambda_c D_s^*$ & 4203.3 & 4202.0 & 0.0 \\
            {} & 3 & $\Xi_c D^*$ & 4591.1 & 4591.5 & $-0.4$ \\
            \hline
            $1(\frac{1}{2})^-$ & 1 & $\Xi_c D$ & 4473.9 & 4472.6 & 0.0 \\
            {} & 2 & $\Xi_c D^*$ & 4592.8 & 4591.5 & 0.0 \\
            {} & 3 & $\Sigma \eta_c$ & 4342.6 & 4340.9 & 0.0 \\
            {} & 4 & $\Sigma J/\psi$ & 4439.5 & 4437.8 & 0.0 \\
            {} & 5 & $\Sigma^* J/\psi$ & 4566.8 & 4565.2 & 0.0 \\
            {} & 6 & $\Sigma_c D_s$ & 4485.7 & 4484.5 & 0.0 \\
            {} & 7 & $\Sigma_c D_s^*$ & 4609.6 & 4608.4 & 0.0 \\
            {} & 8 & $\Sigma_c^* D_s^*$ & 4653.9 & 4652.6 & 0.0 \\
            \hline
            $1(\frac{3}{2})^-$ & 1 & $\Xi_c D^*$ & 4592.8 & 4591.5 & 0.0 \\
            {} & 2 & $\Sigma J/\psi$ & 4439.5 & 4437.8 & 0.0 \\
            {} & 3 & $\Sigma^* \eta_c$ & 4469.9 & 4468.3 & 0.0 \\
            {} & 4 & $\Sigma^* J/\psi$ & 4566.8 & 4565.2 & 0.0 \\
            {} & 5 & $\Sigma_c D_s^*$ & 4609.6 & 4608.4 & 0.0 \\
            {} & 6 & $\Sigma_c^* D_s$ & 4530.0 & 4528.7 & 0.0 \\
            {} & 7 & $\Sigma_c^* D_s^*$ & 4653.9 & 4652.6 & 0.0 \\
            \hline
            $1(\frac{5}{2})^-$ & 1 & $\Sigma^* J/\psi$ & 4566.8 & 4565.2 & 0.0 \\
            {} & 2 & $\Sigma_c^* D_s^*$ & 4653.9 & 4652.6 & 0.0 \\
        \end{tabular}
    \end{ruledtabular}
    \end{table}

    To identify which terms in the Hamiltonian making the state to be bounded, the contributions from each term of Hamiltonian for the pentaquark states, and corresponding baryons and mesons, are given in Table~\ref{tab:energy_component_qqs{bbar}b}. 
$\Delta_E$ is the difference among the contributions to five-quark state and the sum of the corresponding baryon and meson. From Table~\ref{tab:energy_component_qqs{bbar}b} we see there are two types of configurations: 
\begin{enumerate}
\item $(qqs)(\bar{Q}Q)$ configuration, such as $\Lambda \eta_b$, $\Lambda \Upsilon$, $\Sigma^{(*)} \eta_b$ and $\Sigma^{(*)} \Upsilon$ states, and $(qqQ)(\bar{Q}s)$ configuration, such as $\Lambda_b B_s^{(*)}$ and $\Sigma_b^{(*)} B_s^{(*)}$ states. 

Because Goldstone boson exchanges are considered only between light quarks, the contributions of Goldstone boson exchanges in the five-quark state all come from the 3-quark cluster, and there are no Goldstone boson exchanges in the interaction between two clusters. There are also no contributions of confinement potential and one-gluon-exchange because color matrix elements $\langle \chi^c | \boldsymbol{\lambda}_{i}^c \cdot \boldsymbol{\lambda}_{j}^c | \chi^c \rangle  = 0 (i=1,2,3; j=4,5)$ between two clusters. Kinetic energy of relative motion between two clusters provides slight repulsion, with the contribution about $0.6\sim1.4$ MeV. Therefore, it is impossible to form bound states in these channels.

\item $(qsQ)(\bar{Q}q)$ configuration, such as $\Xi_b B^{(*)}$ states. 

Although kinetic energy of relative motion between two clusters still provides major repulsion, considerable contributions of confinement potential and one-gluon-exchange between two clusters appear. Different from the previous situation, due to the exchange of identical particles between two clusters, color matrix elements between two clusters $\langle \chi^c | \boldsymbol{\lambda}_{i}^c \cdot \boldsymbol{\lambda}_{j}^c \mathcal{A} | \chi^c \rangle (i=1,2,3; j=4,5)$ are not generally equal to zero, where $\mathcal{A}$ is the antisymmetrization operator act on color states. The contributions of Goldstone boson exchanges are very small and play a secondary role. Noting that $\sigma$ meson exchange between two clusters always provides attraction different from $\pi$, K, $\eta$ meson exchanges. Therefore, confinement potential, one-gluon-exchange and $\sigma$ meson exchange contribute to the binding of the $\Xi_b B^{(*)}$ states.
\end{enumerate}  

    \begin{table*}
    \caption{\label{tab:energy_component_qqs{bbar}b}The contributions from each term of Hamiltonian in the system of $qqs\bar{b}b$ with quantum numbers $IJ^P = 0(\frac{1}{2})^-$ (unit: MeV).}
    \begin{ruledtabular}
    \begin{tabular}{ c c c c c c c c c c }
            Channel & rest mass & kinetic & $V^C$ & $V^G$ & $V^{\pi}$ & $V^{K}$ & $V^{\eta}$ & $V^{\sigma}$ & total \\
            \hline
            $\Lambda \eta_b$ & 11381.0 & 1585.2 & $-$842.6 & $-$1352.8 & $-$336.7 & 0.0 & 75.0 & $-$26.8 & 10482.3 \\
            $\Lambda$ & 1181.0 & 938.8 & $-$203.9 & $-$614.5 & $-$336.7 & 0.0 & 75.0 & $-$26.8 & 1012.9 \\
            $\eta_b$ & 10200.0 & 645.0 & $-$638.7 & $-$738.3 & 0.0 & 0.0 & 0.0 & 0.0 & 9468.0 \\
            $\Delta_E$ & 0.0 & 1.4 & 0.0 & 0.0 & 0.0 & 0.0 & 0.0 & 0.0 & 1.4 \\
            \hline
            $\Lambda \Upsilon$ & 11381.0 & 1179.5 & $-$823.4 & $-$929.6 & $-$336.7 & 0.0 & 75.0 & $-$26.8 & 10519.0 \\
            $\Lambda$ & 1181.0 & 938.8 & $-$203.9 & $-$614.5 & $-$336.7 & 0.0 & 75.0 & $-$26.8 & 1012.9 \\
            $\Upsilon$ & 10200.0 & 239.3 & $-$619.5 & $-$315.1 & 0.0 & 0.0 & 0.0 & 0.0 & 9504.7 \\
            $\Delta_E$ & 0.0 & 1.4 & 0.0 & 0.0 & 0.0 & 0.0 & 0.0 & 0.0 & 1.4 \\
            \hline
            $\Lambda_b B_s$ & 11381.0 & 1145.6 & $-$597.0 & $-$889.9 & $-$353.5 & 0.0 & 80.8 & $-$25.9 & 10741.1 \\
            $\Lambda_b$ & 5726.0 & 1014.1 & $-$329.3 & $-$727.5 & $-$353.5 & 0.0 & 80.8 & $-$25.9 & 5384.7 \\
            $B_s$ & 5655.0 & 130.9 & $-$267.7 & $-$162.4 & 0.0 & 0.0 & 0.0 & 0.0 & 5355.8 \\
            $\Delta_E$ & 0.0 & 0.6 & 0.0 & 0.0 & 0.0 & 0.0 & 0.0 & 0.0 & 0.6 \\
            \hline
            $\Lambda_b B_s^*$ & 11381.0 & 1138.0 & $-$554.9 & $-$879.7 & $-$353.5 & 0.0 & 80.8 & $-$25.9 & 10785.8 \\
            $\Lambda_b$ & 5726.0 & 1014.1 & $-$329.3 & $-$727.5 & $-$353.5 & 0.0 & 80.8 & $-$25.9 & 5384.7 \\
            $B_s^*$ & 5655.0 & 123.3 & $-$225.6 & $-$152.2 & 0.0 & 0.0 & 0.0 & 0.0 & 5400.5 \\
            $\Delta_E$ & 0.0 & 0.6 & 0.0 & 0.0 & 0.0 & 0.0 & 0.0 & 0.0 & 0.6 \\
            \hline
            $\Xi_b B$ & 11381.0 & 1053.2 & $-$403.6 & $-$879.1 & 8.5 & 0.0 & $-$15.5 & $-$11.6 & 11132.9 \\
            $\Xi_b$ & 5968.0 & 587.3 & $-$255.8 & $-$418.0 & 0.0 & 0.0 & $-$14.5 & 0.0 & 5867.0 \\
            $B$ & 5413.0 & 216.2 & $-$138.3 & $-$213.0 & 0.0 & 0.0 & 0.0 & 0.0 & 5277.9 \\
            $\Delta_E$ & 0.0 & 249.7 & $-$9.5 & $-$248.1 & 8.5 & 0.0 & $-$1.0 & $-$11.6 & $-$12.0 \\
            \hline
            $\Xi_b B^*$ & 11381.0 & 1387.4 & $-$368.3 & $-$1202.1 & 1.6 & 0.0 & $-$13.9 & $-$12.3 & 11173.4 \\
            $\Xi_b$ & 5968.0 & 587.3 & $-$255.8 & $-$418.0 & 0.0 & 0.0 & $-$14.5 & 0.0 & 5867.0 \\
            $B^*$ & 5413.0 & 179.4 & $-$113.7 & $-$159.9 & 0.0 & 0.0 & 0.0 & 0.0 & 5318.8 \\
            $\Delta_E$ & 0.0 & 620.7 & 1.2 & $-$624.2 & 1.6 & 0.0 & 0.6 & $-$12.3 & $-$12.4 \\
        \end{tabular}
    \end{ruledtabular}
    \end{table*}

    The physical pentaquark states must be colorless, but the way of reaching this condition can be acquired through the coupling of two colorless clusters or two colorful clusters. Therefore, there are two kinds of color structures, one is the color singlet-singlet $1 \bigotimes 1$ structure, and another is color octet-octet $8 \bigotimes 8$ structure. With consideration of the color-octet structure, the bound states found in channel coupling calculations of the two kinds of color structures are listed in Table~\ref{tab:boundstatesof_qqs{Qbar}Q}. Where $E_S$ and $E_H$ are the eigen-energy of the color-singlet structure and the color-octet structure, respectively. $E_{S+H}$ is the eigen-energy of the coupling of two kind of color structures. The $E_{th}{\rm (Exp.)}$ is the experimental value of noninteracting baryon-meson threshold. The $E'$ is the predicted pentaquark energies obtained by taking the sum of the experimental baryon-meson threshold and the binding energy.  
    One can see in Table~\ref{tab:boundstatesof_qqs{Qbar}Q} that with the help of the coupling of the color-singlet structure and the color-octet structure, the magnitude of bounding energies for $qqs\bar{b}b$, $qqs\bar{b}c$ and $qqs\bar{c}b$, and $qqs\bar{c}c$ configurations increase about $60\sim110$ MeV, $1\sim12$ MeV, and $1$ MeV, respectively. 
When taking color-octet structure into consideration, the bounding energies for $qqs\bar{b}b$ configuration increase so dramatically, which might be no more molecular states. We will specify this in the following discussions.  

    \begin{table*}
    \caption{\label{tab:boundstatesof_qqs{Qbar}Q}The energies of pentaquark states with the coupling of two color structures (unit: MeV).}
    \begin{ruledtabular}
    \begin{tabular}{ c c c c c c c c c c }
            System & $IJ^p$ & Channel & $E_S$ & $E_H$ & $E_{S+H}$ & $E_{th}$(Theo.) & $E_B$ & $E_{th}$(Exp.) & $E'$ \\
            \hline
            $qqs\bar{b}b$ & $0(\frac{1}{2})^-$ & $\Xi_b B$ & 11132.9 & 11180.5 & 11070.6 & 11144.9 & $-$74.3 & 11071.2 & 10996.9 \\
            {} & $0(\frac{1}{2})^-$ & $\Xi_b B^*$ & 11173.4 & 11171.9 & 11062.8 & 11185.8 & $-$123.0 & 11116.5
 & 10993.5
 \\ 
            {} & $0(\frac{3}{2})^-$ & $\Xi_b B^*$ & 11175.3 & 11186.5 & 11078.7 & 11185.8 & $-$107.1 & 11116.5 & 11009.4 \\
            \hline
            $qqs\bar{b}c$ & $0(\frac{1}{2})^-$ & $\Xi_c B$ & 7846.9 & 7971.4 & 7844.7 & 7852.1 & $-$7.4 & 7747.2 & 7739.8 \\
            {} & $0(\frac{1}{2})^-$ & $\Xi_c B^*$ & 7887.4 & 7958.7 & 7881.3 & 7893.0 & $-$11.7 & 7792.5 & 7780.8 \\ 
            {} & $0(\frac{3}{2})^-$ & $\Xi_c B^*$ & 7888.7 & 7979.8 & 7883.7 & 7893.0 & $-$9.3 & 7792.5 & 7783.2 \\
            \hline
            $qqs\bar{c}b$ & $0(\frac{1}{2})^-$ & $\Xi_b D$ & 7760.7 & 7963.9 & 7759.6 & 7765.4 & $-$5.8 & 7661.5 & 7655.7 \\
            {} & $0(\frac{1}{2})^-$ & $\Xi_b D^*$ & 7880.8 & 7954.9 & 7868.9 & 7884.3 & $-15.4$ & 7798.7 & 7783.3 \\ 
            {} & $0(\frac{3}{2})^-$ & $\Xi_b D^*$ & 7881.7 & 7972.6 & 7876.2 & 7884.3 & $-$8.1 & 7798.7 & 7790.6 \\
            \hline
            $qqs\bar{c}c$ & $0(\frac{1}{2})^-$ & $\Xi_c D$ & 4471.4 & 4706.5 & 4470.9 & 4472.6 & $-$1.7 & 4337.5 & 4335.8 \\
            {} & $0(\frac{1}{2})^-$ & $\Xi_c D^*$ & 4590.7 & 4688.1 & 4589.9 & 4591.5 & $-$1.6 & 4474.7 & 4473.1 \\ 
            {} & $0(\frac{3}{2})^-$ & $\Xi_c D^*$ & 4591.1 & 4721.1 & 4590.5 & 4591.5 & $-$1.0 & 4474.7 & 4473.7 \\
    \end{tabular}
    \end{ruledtabular}
    \end{table*}

    The spacial configurations of the states are determined by the dynamical calculation. The root-mean-square (RMS) distances between any two quarks for the bound states in color-singlet structure  and the coupling of two color structures are calculated and shown in Tables~\ref{tab:RMSof_qqs{Qbar}Q_1} and \ref{tab:RMSof_qqs{Qbar}Q_18}, respectively. 
    From Tables~\ref{tab:RMSof_qqs{Qbar}Q_1} and \ref{tab:RMSof_qqs{Qbar}Q_18} we see that:
\begin{enumerate}
\item For color-singlet structure, the distances among the 3-quark cluster range from $0.5$ to $0.6$ fm,  
while the distances between two clusters are around $0.9-5.4$ fm. The result indicates that they may be molecular states. 
\item For the coupling of two color structures with $qqs\bar{b}b$ configuration, $r_{34}$ is about $0.32\sim0.33$ fm, much smaller than all other distances, this implies that there is a compact $\bar{b}b$-pair surrounded by three other quarks.
\item For the coupling of two color structures  with $qqs\bar{b}c$, $qqs\bar{c}b$ and $qqs\bar{c}c$ configurations, the distances among the 3-quark cluster range from $0.5$ to $0.8$ fm, 
while the distances between two clusters are around $0.5-4.1$ fm. The result indicates that they may be compact molecular states. 
\end{enumerate}

    \begin{table*}
    \caption{\label{tab:RMSof_qqs{Qbar}Q_1}The RMS distances between any two quarks with color-singlet structure (unit: fm).}
    \begin{ruledtabular}
    \begin{tabular}{ c c c c c c c c c c c c c c }
            System & $IJ^p$ & Channel &  $N_{123,45}$ & $r_{12}$ & $r_{13}$ & $r_{23}$ & $r_{45}$ & $r_{14}$ & $r_{15}$ & $r_{24}$ & $r_{25}$ & $r_{34}$ & $r_{35}$ \\
            \hline
            $qqs\bar{b}b$ & $0(\frac{1}{2})^-$ & $\Xi_b B$ & $qsb,\bar{b}q$ & 0.88 & 0.81 & 0.53 & 0.78 & 0.78 & 1.11 & 0.98 & 0.88 & 0.86 & 0.81 \\
            {} & $0(\frac{1}{2})^-$ & $\Xi_b B^*$ & $qsb,\bar{b}q$ & 0.91 & 0.83 & 0.54 & 0.80 & 0.80 & 1.14 & 1.03 & 0.91 & 0.90 & 0.83 \\
            {} & $0(\frac{3}{2})^-$ & $\Xi_b B^*$ & $qsb,\bar{b}q$ & 0.91 & 0.83 & 0.54 & 0.80 & 0.80 & 1.15 & 1.04 & 0.91 & 0.91 & 0.83 \\
            \hline
            $qqs\bar{b}c$ & $0(\frac{1}{2})^-$ & $\Xi_c B$ & $qsc,\bar{b}q$ & 0.94 & 0.88 & 0.61 & 0.82 & 0.82 & 1.16 & 1.04 & 0.94 & 0.94 & 0.88 \\
            {} & $0(\frac{1}{2})^-$ & $\Xi_c B^*$ & $qsc,\bar{b}q$ & 0.96 & 0.91 & 0.61 & 0.84 & 0.84 & 1.20 & 1.08 & 0.96 & 0.99 & 0.91 \\
            {} & $0(\frac{3}{2})^-$ & $\Xi_c B^*$ & $qsc,\bar{b}q$ & 0.99 & 0.93 & 0.61 & 0.86 & 0.86 & 1.23 & 1.12 & 0.99 & 1.02 & 0.93 \\
            \hline
            $qqs\bar{c}b$ & $0(\frac{1}{2})^-$ & $\Xi_b D$ & $qsb,\bar{c}q$ & 0.96 & 0.88 & 0.54 & 0.86 & 0.86 & 1.21 & 1.12 & 0.96 & 1.00 & 0.88 \\
            {} & $0(\frac{1}{2})^-$ & $\Xi_b D^*$ & $qsb,\bar{c}q$ & 1.25 & 1.19 & 0.54 & 1.17 & 1.17 & 1.65 & 1.60 & 1.25 & 1.51 & 1.19 \\
            {} & $0(\frac{3}{2})^-$ & $\Xi_b D^*$ & $qsb,\bar{c}q$ & 1.28 & 1.22 & 0.54 & 1.21 & 1.21 & 1.71 & 1.65 & 1.28 & 1.57 & 1.22 \\
            \hline
            $qqs\bar{c}c$ & $0(\frac{1}{2})^-$ & $\Xi_c D$ & $qsc,\bar{c}q$ & 2.48 & 2.46 & 0.61 & 2.44 & 2.44 & 3.45 & 3.42 & 2.48 & 3.39 & 2.46 \\
            {} & $0(\frac{1}{2})^-$ & $\Xi_c D^*$ & $qsc,\bar{c}q$ & 3.67 & 3.66 & 0.61 & 3.65 & 3.65 & 5.16 & 5.14 & 3.67 & 5.12 & 3.66 \\
            {} & $0(\frac{3}{2})^-$ & $\Xi_c D^*$ & $qsc,\bar{c}q$ & 3.85 & 3.83 & 0.61 & 3.82 & 3.82 & 5.40 & 5.38 & 3.85 & 5.36 & 3.83 \\
    \end{tabular}
    \end{ruledtabular}
    \end{table*}

    \begin{table*}
    \caption{\label{tab:RMSof_qqs{Qbar}Q_18}The RMS distances between any two quarks with the coupling of two color structures (unit: fm).}
    \begin{ruledtabular}
    \begin{tabular}{ c c c c c c c c c c c c c c }
            System & $IJ^p$ & Channel &  $N_{123,45}$ & $r_{12}$ & $r_{13}$ & $r_{23}$ & $r_{45}$ & $r_{14}$ & $r_{15}$ & $r_{24}$ & $r_{25}$ & $r_{34}$ & $r_{35}$ \\
            \hline
            $qqs\bar{b}b$ & $0(\frac{1}{2})^-$ & $\Xi_b B$ & $qsb,\bar{b}q$ & 0.95 & 0.80 & 0.59 & 0.80 & 0.80 & 1.10 & 0.61 & 0.95 & 0.33 & 0.80 \\
            {} & $0(\frac{1}{2})^-$ & $\Xi_b B^*$ & $qsb,\bar{b}q$ & 0.94 & 0.79 & 0.59 & 0.79 & 0.79 & 1.09 & 0.61 & 0.94 & 0.32 & 0.79 \\
            {} & $0(\frac{3}{2})^-$ & $\Xi_b B^*$ & $qsb,\bar{b}q$ & 0.95 & 0.81 & 0.60 & 0.80 & 0.80 & 1.11 & 0.62 & 0.95 & 0.33 & 0.81 \\
            \hline
            $qqs\bar{b}c$ & $0(\frac{1}{2})^-$ & $\Xi_c B$ & $qsc,\bar{b}q$ & 0.91 & 0.84 & 0.62 & 0.75 & 0.75 & 1.08 & 0.90 & 0.91 & 0.78 & 0.84 \\
            {} & $0(\frac{1}{2})^-$ & $\Xi_c B^*$ & $qsc,\bar{b}q$ & 0.95 & 0.85 & 0.69 & 0.77 & 0.77 & 1.08 & 0.75 & 0.95 & 0.59 & 0.85 \\
            {} & $0(\frac{3}{2})^-$ & $\Xi_c B^*$ & $qsc,\bar{b}q$ & 0.93 & 0.84 & 0.66 & 0.76 & 0.76 & 1.08 & 0.80 & 0.93 & 0.66 & 0.84 \\
            \hline
            $qqs\bar{c}b$ & $0(\frac{1}{2})^-$ & $\Xi_b D$ & $qsb,\bar{c}q$ & 0.92 & 0.83 & 0.54 & 0.81 & 0.81 & 1.14 & 1.03 & 0.92 & 0.90 & 0.83 \\
            {} & $0(\frac{1}{2})^-$ & $\Xi_b D^*$ & $qsb,\bar{c}q$ & 0.98 & 0.83 & 0.62 & 0.86 & 0.86 & 1.15 & 0.72 & 0.98 & 0.49 & 0.83 \\
            {} & $0(\frac{3}{2})^-$ & $\Xi_b D^*$ & $qsb,\bar{c}q$ & 1.16 & 1.03 & 0.82 & 1.06 & 1.06 & 1.39 & 0.92 & 1.16 & 0.63 & 1.03 \\
            \hline
            $qqs\bar{c}c$ & $0(\frac{1}{2})^-$ & $\Xi_c D$ & $qsc,\bar{c}q$ & 2.03 & 2.00 & 0.61 & 1.98 & 1.98 & 2.80 & 2.76 & 2.03 & 2.72 & 2.00 \\
            {} & $0(\frac{1}{2})^-$ & $\Xi_c D^*$ & $qsc,\bar{c}q$ & 2.54 & 2.52 & 0.61 & 2.50 & 2.50 & 3.53 & 3.49 & 2.54 & 3.46 & 2.52 \\
            {} & $0(\frac{3}{2})^-$ & $\Xi_c D^*$ & $qsc,\bar{c}q$ & 2.94 & 2.92 & 0.61 & 2.91 & 2.91 & 4.11 & 4.08 & 2.94 & 4.05 & 2.92 \\
    \end{tabular}
    \end{ruledtabular}
    \end{table*}

\section{\label{sec:level4}Summary}
    In this work, we investigate the five-quark system of $qqs\bar{Q}Q$ configuration, with all possible quantum numbers $IJ^P = 0(\frac{1}{2})^-$, $IJ^P = 0(\frac{3}{2})^-$, $IJ^P = 1(\frac{1}{2})^-$, $IJ^P = 1(\frac{3}{2})^-$, $IJ^P = 1(\frac{5}{2})^-$, in the framework of the chiral quark model. 
    
    For isospin $I = 0$ in the system of $qqs\bar{b}b$, several pentaquark bound states $\Xi_b B(J = 1/2)$, $\Xi_b B^*(J = 1/2)$ and $\Xi_b B^*(J = 3/2)$, have been predicted to exist in the color-singlet structure. For isospin $I = 1$, we find no bound states in all possible channels. We replace some b-quark by c-quark, and search for possible bound states systematically in systems of $qqs\bar{b}c$, $qqs\bar{c}b$ and $qqs\bar{c}c$. The similar results are obtained compared with the system of $qqs\bar{b}b$. 
    Then we add the color-octet structure into the calculation.     
    The result indicates that: (1), taking color-octet structure into consideration always provides more bounding energy than color-singlet structure; (2), the more heavier quark prevents, the easier to form the bound states. 
    
    The distances between any two quarks for the bound states are calculated, and the result suggests that: (1), $qqs\bar{b}b$ configuration changes the structure from molecular state to a compact $\bar{b}b$-pair surrounded by three other quarks, when color-octet structure is considered; (2), $qqs\bar{b}c$, $qqs\bar{c}b$ and $qqs\bar{c}c$ configurations are always molecular states in the present calculation. 

    Finally, we expect that relevant collaborations will make an attempt to search for pentaquark states with heavy flavors in the future.
\begin{acknowledgments}
B.R.~He was supported in part by the National Natural Science Foundation of China (Grant No. 11705094), Natural Science Foundation of Jiangsu Province, China (Grant No. BK20171027), Natural Science Foundation of the Higher Education Institutions of Jiangsu Province, China (Grant No. 17KJB140011), and by the Research Start-up Funding (B.R.~He) of Nanjing Normal University. 
And the work of J.L.~Ping was supported in part by the National Science Foundation of China under Grants No. 11775118, and No. 11535005. 
\end{acknowledgments}


\begin{thebibliography}{99}


\bibitem{GellMann:1964nj}
M.~Gell-Mann,
Phys. Lett. \textbf{8}, 214-215 (1964).


\bibitem{Zweig:1964jf}
G.~Zweig,
CERN-TH-412.


\bibitem{Nakamura:2010zzi}
K.~Nakamura \textit{et al.} [Particle Data Group],
J. Phys. G \textbf{37}, 075021 (2010).


\bibitem{Guo:2017jvc}
F.~Guo, C.~Hanhart, U.~Meißner, Q.~Wang, Q.~Zhao and B.~Zou,
Rev. Mod. Phys. \textbf{90}, 015004 (2018).




\bibitem{Kaiser:1995cy}
N.~Kaiser, P.~Siegel and W.~Weise,
Phys. Lett. B \textbf{362}, 23-28 (1995).


\bibitem{Oller:2000fj}
J.~Oller and U.~G.~Meissner,
Phys. Lett. B \textbf{500}, 263-272 (2001). 


\bibitem{Inoue:2001ip}
T.~Inoue, E.~Oset and M.~Vicente Vacas,
Phys. Rev. C \textbf{65}, 035204 (2002). 


\bibitem{GarciaRecio:2003ks}
C.~Garcia-Recio, M.~Lutz and J.~Nieves,
Phys. Lett. B \textbf{582}, 49-54 (2004). 


\bibitem{Helminen:2000jb}
C.~Helminen and D.~Riska,
Nucl. Phys. A \textbf{699}, 624-648 (2002).

\bibitem{Zhu:2004xa}
S.~L.~Zhu,
Int. J. Mod. Phys. A \textbf{19}, 3439-3469 (2004). 


\bibitem{Liu:2005pm}
B.~Liu and B.~Zou,
Phys. Rev. Lett. \textbf{96}, 042002 (2006).


\bibitem{Bijker:2009up}
R.~Bijker, E.~Santopinto and E.~Santopinto,
Phys. Rev. C \textbf{80}, 065210 (2009).


\bibitem{An:2009uv}
C.~An and B.~Zou,
Sci. China G \textbf{52}, 1452-1457 (2009).

\bibitem{Liu:2019zoy}
Y.~R.~Liu, H.~X.~Chen, W.~Chen, X.~Liu and S.~L.~Zhu,
Prog. Part. Nucl. Phys. \textbf{107}, 237-320 (2019). 


\bibitem{Aaij:2015tga}
R.~Aaij \textit{et al.} [LHCb],
Phys. Rev. Lett. \textbf{115}, 072001 (2015).


\bibitem{Aaij:2019vzc}
R.~Aaij \textit{et al.} [LHCb],
Phys. Rev. Lett. \textbf{122},  222001 (2019).

\bibitem{He:2015cea}
J.~He,
Phys. Lett. B \textbf{753}, 547-551 (2016). 

\bibitem{Liu:2015fea}
X.~H.~Liu, Q.~Wang and Q.~Zhao,
Phys. Lett. B \textbf{757}, 231-236 (2016). 

\bibitem{Scoccola:2015nia}
N.~Scoccola, D.~Riska and M.~Rho,
Phys. Rev. D \textbf{92}, 051501 (2015). 

\bibitem{Wang:2015ava}
Z.~G.~Wang and T.~Huang,
Eur. Phys. J. C \textbf{76}, 43 (2016). 

\bibitem{Shen:2016tzq}
C.~W.~Shen, F.~K.~Guo, J.~J.~Xie and B.~S.~Zou,
Nucl. Phys. A \textbf{954}, 393-405 (2016). 

\bibitem{Zhou:2018bkn}
Q.~S.~Zhou, K.~Chen, X.~Liu, Y.~R.~Liu and S.~L.~Zhu,
Phys. Rev. C \textbf{98}, 045204 (2018). 

\bibitem{Li:2018vhp}
S.~Y.~Li, Y.~R.~Liu, Y.~N.~Liu, Z.~G.~Si and J.~Wu,
Eur. Phys. J. C \textbf{79}, 87 (2019).

\bibitem{An:2019idk}
H.~T.~An, Q.~S.~Zhou, Z.~W.~Liu, Y.~R.~Liu and X.~Liu,
Phys. Rev. D \textbf{100}, 056004 (2019). 

\bibitem{Zhang:2020erj}
B.~T.~Zhang, J.~S.~Wang and Y.~L.~Ma,
[arXiv:2002.10954 [hep-ph]].


\bibitem{Wu:2010jy}
J.~Wu, R.~Molina, E.~Oset and B.~Zou,
Phys. Rev. Lett. \textbf{105}, 232001 (2010).


\bibitem{Wu:2010rv}
J.~Wu and B.~Zou,
Phys. Lett. B \textbf{709}, 70-76 (2012).


\bibitem{Yang:2018oqd}
G.~Yang, J.~Ping and J.~Segovia,
Phys. Rev. D \textbf{99}, 014035 (2019).


\bibitem{1797080}
X.~Liu, H.~Huang and J.~Ping,
[arXiv:2005.09646 [hep-ph]].



\bibitem{Shimizu:2016rrd}
Y.~Shimizu, D.~Suenaga and M.~Harada,
Phys. Rev. D \textbf{93}, 114003 (2016).


\bibitem{Aaij:2017jgf}
R.~Aaij \textit{et al.} [LHCb],
Phys. Rev. D \textbf{97}, 032010 (2018).


\bibitem{Obukhovsky:1990tx}
I.~Obukhovsky and A.~Kusainov,
Phys. Lett. B \textbf{238}, 142-148 (1990). 

\bibitem{Fernandez:1993hx}
F.~Fernandez, A.~Valcarce, U.~Straub and A.~Faessler,
J. Phys. G \textbf{19}, 2013-2026 (1993). 

\bibitem{Yu:1995ag}
Y.~Yu, Z.~Zhang, P.~Shen and L.~Dai,
Phys. Rev. C \textbf{52}, 3393-3398 (1995). 

\bibitem{Valcarce:1995dm}
A.~Valcarce, F.~Fernandez, P.~Gonzalez and V.~Vento,
Phys. Lett. B \textbf{367}, 35-39 (1996). 

\bibitem{Fernandez:2019ses}
F.~Fernández, P.~G.~Ortega and D.~R.~Entem,
Front. in Phys. \textbf{7}, 233 (2020)

\bibitem{Vijande:2004he}
J.~Vijande, F.~Fernandez and A.~Valcarce,
J. Phys. G \textbf{31}, 481 (2005).



\bibitem{Segovia:2013wma}
J.~Segovia, D.~Entem, F.~Fernandez and E.~Hernandez,
Int. J. Mod. Phys. E \textbf{22}, 1330026 (2013).


\bibitem{Hiyama:2003cu}
E.~Hiyama, Y.~Kino and M.~Kamimura,
Prog. Part. Nucl. Phys. \textbf{51}, 223-307 (2003).














\bibitem{Park:2016xrw}
A.~Park, P.~Gubler, M.~Harada, S.~H.~Lee, C.~Nonaka and W.~Park,
Phys. Rev. D \textbf{93}, 054035 (2016). 



\end{thebibliography}
\end{document}